\documentclass[%
superscriptaddress,
twocolumn,
amsmath,amssymb,
aps,
pra
]{revtex4-1}

\usepackage{dcolumn}
\usepackage{ctable}
\usepackage{graphicx}
\usepackage{dcolumn}
\usepackage{bm}
\usepackage[utf8]{inputenc}
\usepackage{xcolor}
\usepackage{soul}
\usepackage{hyperref}

\begin{document}

\title{Strong coupling of a single photon to a magnetic vortex}

\author{María José Martínez-Pérez}
 \email{pemar@unizar.es}
\affiliation {Instituto de Ciencia de Materiales de
  Aragón and Departamento de Física de la Materia Condensada ,
  CSIC-Universidad de Zaragoza, Pedro Cerbuna 12, 50009 Zaragoza,
  Spain}
 
  \affiliation{Fundación ARAID, Avda. de Ranillas, 50018 Zaragoza, Spain}
\author{David Zueco}
\email{dzueco@unizar.es}
\affiliation {Instituto de Ciencia de Materiales de
  Aragón and Departamento de Física de la Materia Condensada ,
  CSIC-Universidad de Zaragoza, Pedro Cerbuna 12, 50009 Zaragoza,
  Spain}
\affiliation{Fundación ARAID, Avda. de Ranillas, 50018 Zaragoza, Spain}

\begin{abstract}
Strong light-matter coupling means that  cavity photons  and other type of  matter excitations are coherently exchanged. It is used to couple different qubits (matter) via a quantum bus (photons) or to communicate different type of excitations, \textit{e.g.}, transducing light into  phonons or magnons. An unexplored, so far, interface is the coupling between light and topologically protected particle-like excitations as magnetic domain walls, skyrmions or vortices.  Here, we show theoretically that a single photon living in a superconducting cavity can be strongly coupled  to the gyrotropic mode of a magnetic vortex in a nanodisc.  We combine numerical and analytical calculations for a superconducting coplanar waveguide resonator and different realizations of the nanodisc (materials and sizes).  We show that, for enhancing the coupling, constrictions fabricated in the resonator are crucial, allowing to reach the strong coupling in CoFe discs of radius $200-400$ nm having  resonance frequencies of few GHz.
The strong coupling regime permits to coherently exchange  a single photon and quanta of vortex gyration.  Thus, our calculations show that the device proposed here serves as a transducer between photons and gyrating vortices, opening the way to complement superconducting qubits with topologically protected spin-excitations like vortices or skyrmions. We finish by discussing potential applications in quantum data processing based on the exploitation of the vortex as a short-wavelength magnon emitter.
\end{abstract}

\maketitle

\section{Introduction}

Nanoscopic magnetic systems are a natural playground for the nucleation, manipulation and study of topological particle-like solitons.\cite{Braun12}
For instance, a domain wall in an easy-axis magnet constitutes a two-dimensional topological soliton. 
Pure easy-plane spins in three dimensions can be arranged to contain a one-dimensional topological defect in the form of a vortex line. 
Finally, isotropic spins can also exhibit a zero-dimension topological defect named Bloch point. 
%
%
Skyrmions constitute a different kind of topologically protected defect where the magnetization order parameter does not vanish.\cite{Bogdanov06}
Being topological, these particle-like objects are extremely stable against thermal fluctuations or material defects.
Additionally, domain walls and skyrmions can be efficiently moved using low-power spin currents or alternating fields which makes them very attractive for their integration into the well-known racetrack magnetic memory.\cite{Fert13,Parkin10}

Magnetic vortices are stabilized in thin film confined geometries.\cite{Shinjo00}
Here, the magnetization curls clockwise or counter-clockwise (defining the vortex circulation: C) in order to minimize the total magnetostatic energy. 
In the core, exchange interaction forces the magnetization to point up or down (defining the vortex polarity: P) leading to an extremely stable four-state logic unit.\cite{Pigeau10}
The vortex core gyrates around its equilibrium position at sub-GHz and GHz frequencies. 
This gyrotropic mode can be used to promote polarity inversion when sufficiently large vortex velocities are reached, similarly to the well-known process of walker breakdown in domain walls.\cite{Waeyenberge06,Keisuke07}.
Interestingly, static spin-polarized currents can be used to produce self-sustained vortex gyration leading to the development of spin-torque vortex nanoscillators.\cite{Pribiag07}
Vortex gyration can also be used to generate short-wavelength (sub-micrometric) incoherent\cite{Lee05} or coherent spin-waves.\cite{Wintz16}

Spin-waves (or their corresponding quasiparticles, magnons) carry information on the spin angular momentum at GHz and sub-THz frequencies with negligible heat dissipation. 
Due to the possibility of propagating short wavelength exchange spin-waves, 
magnon-based devices could be miniaturized well down to the sub-10 nm scale being very attractive for novel information technologies such as data processing and computing wave-based architectures.\cite{Chumak15} 
Since recently, the interest in magnonic-devices has also moved towards the field of light-matter interaction for quantum information and processing.
Mastering hybrid magnon-photon states opens the way for, \textit{e.g}., coherently coupling distant spin-waves between them\cite{Lambert16,Rameshti18} or to different solid-state realizations of qubits.\cite{Tabuchi15}
So far, most theoretical\cite{SoykalFlatte10} and experimental studies\cite{ZahngNpj15} have focused on small YIG Yttrium--iron--garnet spheres coupled to the quasi-uniform magnetic field region in 3D cavities, so to only excite the uniform Kittel mode (having infinite wavelength).
Magnon-photon coupling in the quantum regime using superconducting coplanar waveguide (CPW) resonators has been addressed only rarely.\cite{Huebl13,Morris17}
Also very few works deal with magnetostatic (long-wavelength) spin-wave modes arising in confined geometries\cite{Zhang16,Bourhill16,Osada2018} and, notably, individual magnetic solitons such as vortices in soft-magnetic discs {\color{black} still need to be explored in detail.\cite{Graf18}}

Here, our mission is to figure out if it is possible to coherently couple   microwave photons and vortex {\color{black} motion}  in the quantum regime. 
In doing so, we open the possibility to communicate two quantum buses: photons, that are easily coupled to, $\textit{e.g.}$, superconducting qubits, and  a topological particle-like soliton that can carry information but can also be used to emit and manipulate sub-micron spin-waves. 
{\color{black}
It is important to notice that, in this work, topology stabilizes the vortex solution but not its gyrotropic motion, which is limited by  the damping in the material. 
}
Below, we show theoretically that it is indeed possible to strongly couple a single photon to the gyrotropic motion of a vortex in a disc of CoFe.    
{\color{black}
Thanks to the fact that the vortex resonance frequency can be modulated by means of an external DC magnetic field,
the coupling can be switched on/off in the few $ns$-regime using standard electronics.
} 
Thus, we obtain  a tunable device.  Our simulations take realistic material parameters, both for the CPW resonator and the magnetic disc, making our proposal feasible in the lab.
%

\section{Results}

\begin{figure*}[t]
\includegraphics[width=0.8\textwidth]{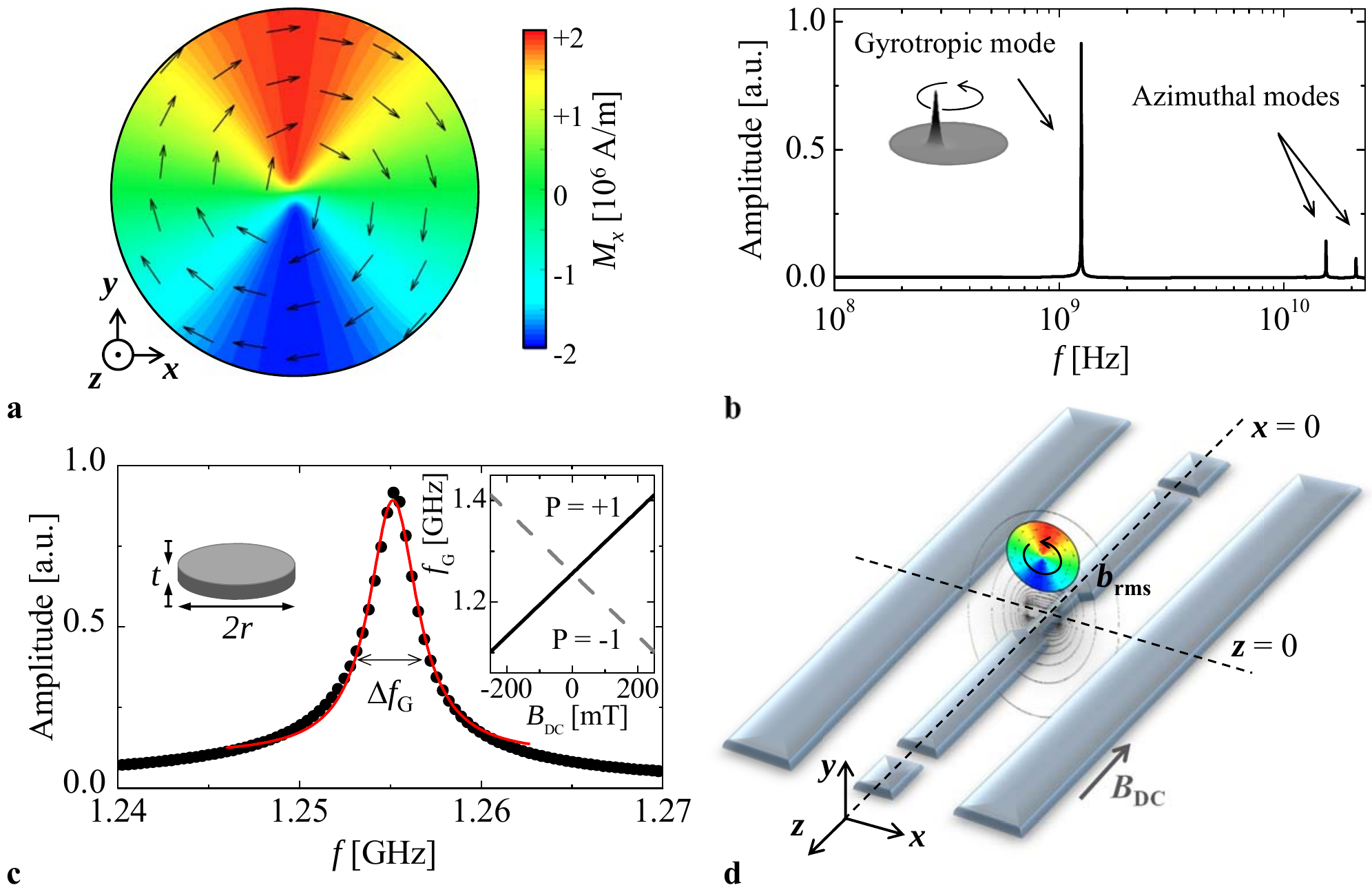}
\caption{\textbf{a}: Numerically calculated spatial distribution of the $M_x$ component for a \textit{C}=+1, \textit{P}=+1 magnetic vortex stabilized in a nanodisc with $r = 200$ nm and $t = 30$ nm. Arrows indicate the in-plane curling of the magnetization. \textbf{b}: Numerically calculated first resonant modes peculiar to the nanodisc shown in \textbf{a}. \textbf{c}: Enlarged view of the gyrotropic mode. Solid line is a fitting to a  lorentzian function. {\color{black}The left inset serves to define the nanodisc dimensions.} The right inset shows the linear dependence of $f_{\rm G}$ on the out-of-plane applied magnetic field $B_{\rm DC}$ for the $P=\pm 1$ vortex. \textbf{d}: Schematic representation of the proposed experiment. The nanodisc lies perpendicular to the central conductor of a superconducting CPW resonator so that the microwave magnetic field $b_{\rm rms}$ (externally applied field $B_{\rm DC}$) is applied parallel (perpendicular) to the nanodisc plane. The coordinate axes $(x,y,z)$ are defined as well with $y=0$ corresponding to the top part of the central conductor.
}
\label{Fig1}
\end{figure*}

\subsection{Vortex description}
 {\color{black} Vortices living in ferromagnetic nanodiscs exhibit a number of resonant  modes. Among them, the gyrotropic mode corresponds to the vortex precessing at frequency $f_{\rm G}$ around its equilibrium position and being only minimally distorted. The gyration sense, on the other hand, is solely given by the right-hand rule to the vortex polarity.
For discs having small aspect ratio, \textit{i.e.}, $t/r \ll 1$, this frequency is approximately proportional to $f_{\rm G} \propto M_{\rm s} t/r$,\cite{Guslienko2006PRL} with $M_{\rm s}$ the material's saturation magnetization, $r$ the radius and $t$ the thickness of the disc (see left inset in Fig. \ref{Fig1}\textbf{c}).
Our purpose is to couple this mode to microwave superconducting resonators.
For this reason, it will be convenient to use nanodiscs having large $t/r$ aspect ratio made out of ferromagnetic materials with large $\mu_0 M_{\rm s} \sim 1$ T so that $f_{\rm G}$ is well in the GHz range.
We highlight that the resonant frequency  can not be made arbitrary large since $f_{\rm G}$ decreases when the thickness reaches several tens of nm. The later can be appreciated in Table \ref{tablenew}. In addition to that, increasing too much the aspect ratio will tend to stabilize a (quasi) uniform magnetization state instead of the vortex.}
%

\begin{table}
\centering
\caption{Numerically calculated values of the gyrotropic frequency ($f_{\rm G}$), line-width ($\Delta f_{\rm G}$) and damping parameter [$\alpha_{\rm v}$, defined in Eq. \eqref{linewidth}] for discs with different radius ($r$) and thicknesses ($t$).}
\label{tablenew}
\begingroup
\setlength{\tabcolsep}{10pt} 
\def\arraystretch{1.4}
\begin{tabular}{ccccc}

$r$ & $t$ & $f_{\rm G}$ & $\Delta f_{\rm G}$ & $\alpha_{\rm v}$  \\
$[$nm$]$  & [nm]  &  [GHz]        &  [MHz]   & [$\times 10^{-3}$] \\  
\hline
100  & 15  &  1.402        &   3.5    &  5.0     \\
200  & 30  &  1.255        &   3.3    &  5.3     \\
400  & 60  &  1.093        &   3.0    &  5.5     \\
\hline
\end{tabular}
\endgroup
\end{table}

The microscopic magnetic structure of a typical vortex state in a magnetic nanodisc with $r = 200$ nm and $t = 30$ nm is shown in Fig. \ref{Fig1}\textbf{a}.
It has been estimated by means of micromagnetic simulations (see \textbf{Methods} section).
This configuration corresponds to a clockwise (\textit{C}=+1) in-plane curling and a central vortex pointing along the positive $z$ direction (\textit{P}=+1).
We highlight that, the discussion made here is independent of the vortex circulation or polarity, that can be set as initial conditions.
We choose a low-damping ferromagnet such as Co$_x$Fe$_{1-x}$ alloy 
(see \textbf{Supporting Information} for a more detailed discussion on the material choice). 
Being metallic, this material exhibits record low values of the Gilbert damping parameter $\alpha_{\rm LLG} \sim 5 \times 10^{-4} $ for $x=0.25$. {\color{black} Conveniently, Co$_x$Fe$_{1-x}$ alloys also offer a relatively large saturation magnetization of $\mu_0 M_{\rm s} \sim 2.4$ T for $x=0.25$.\cite{Schoen16}}
%
%
%

The excitation spectrum  peculiar to the vortex can be simulated by applying a time-dependent homogeneous in-plane perturbation field along the $x$ direction: $b_x(\tau) = A{\rm sinc}(2\pi f_{\rm cutoff} \, \tau)$.
%
%
Such perturbation is equivalent to exciting all eigenmodes susceptible to in-plane magnetic fields at frequencies below $ f_{\rm cutoff}= 50$ GHz.
We obtain the resulting time-dependent spatially-averaged magnetization projected along the $x$ direction $M_x(\tau)$ over a total time $\tau =3$ $\mu$s.
Calculating the corresponding fast Fourier transform (FFT) results in the excitation spectrum shown in Fig. \ref{Fig1}\textbf{b}.
The first mode obtained at $f_{\rm G}=1.255$ GHz is the previously described gyrotropic mode.
%
%
%
%
%
The second and third modes, visible at higher frequencies of $15.41$ and $20.94$ GHz, respectively, are high-order azimuthal modes {\color{black}that will not be discussed here}. 
%

%
The line-width of the gyrotropic mode can be characterized by fitting the corresponding resonance to a lorentzian function giving $\Delta f_{\rm G} = 3.3$ MHz (see Fig. \ref{Fig1}\textbf{c}).
Interestingly, applying an uniform out-of-plane magnetic field ($B_{\rm DC}$) modifies slightly the vortex magnetization profile leading to a linear dependence of $f_{\rm G} $ on $B_{\rm DC}$.\cite{deLoubens2009}
This is exemplified in the right inset of Fig. \ref{Fig1}\textbf{c} for $B_{\rm DC}$ applied along the positive $z$ direction.
Details of our numerical simulations are given in the {\bf Methods} section.

The vortex can, therefore, be treated as a harmonic oscillator driven by a time-dependent in-plane magnetic field $b_{x}(\tau) = b^x_{\rm mw} \cos (2\pi f \tau)$.\cite{Krger2007,Krger2008}
 At resonance, \textit{i.e.}, $f = f_{\rm G}$, the amplitude of the resulting oscillating magnetization will be maximum. 
In the linear (low amplitude) regime, this is given by the vortex susceptibility and the driving amplitude at the disc center position ($\textbf{r}_{\rm c}$): $\Delta M_x = \chi_x(f_{\rm G})  b^x_{\rm mw}(\textbf{r}_{\rm c})$.
Here, $\Delta M_x$ represents the maximum of the spatially-averaged magnetization along the $x$ direction.
%

\begin{figure*}[t]
\includegraphics[width=0.99\textwidth]{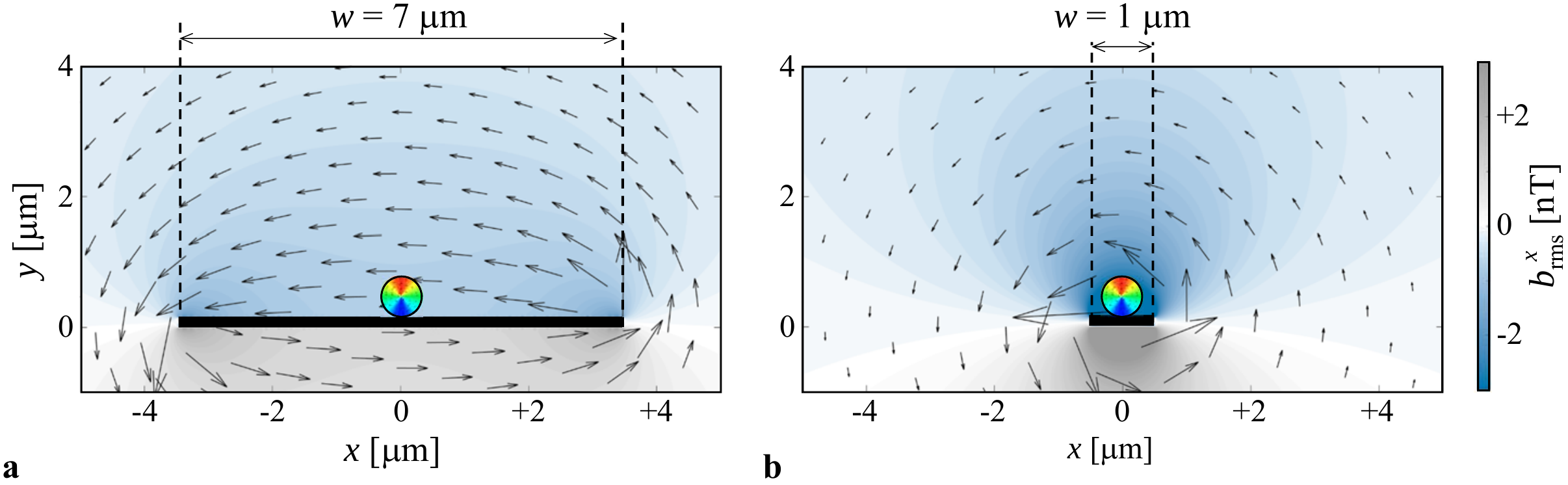}
\caption{Numerically calculated spatial distribution of the amplitude of the $x$ component of the rms magnetic field ($b^x_{\rm rms}$) at the midpoint of the central conductor, \textit{i.e.}, $z=0$. Arrows represent the $(x,y)$-component of $b_{\rm rms}$ with size proportional to the total in-plane amplitude. $w$ is the width of the central Nb conductor (black rectangle) being $150$ nm-thick. \textbf{a}: $w$=7 $\mu$m.  \textbf{b}: $w$=1 $\mu$m.
}
\label{Fig2}
\end{figure*}

\subsection{Vortex-CPW coupling}
Coherent coupling between the vortex gyrotropic mode and resonant photons will be analyzed in the framework of superconducting CPW resonators.
%
Such devices can routinely reach quality factors $Q \sim 10^4 - 10^5$ and operate
under large static in-plane external magnetic fields of few 100 mT.
Strong coupling between the vacuum fluctuations of the CPW and superconducting qubits\cite{Wallraff2004} inaugurated the field of circuit Quantum ElectroDynamics (QED) allowing to process quantum information with superconducting circuits.\cite{Buluta2011} Photons in the resonator may carry the information shared between different qubits or may be used to readout the qubit(s) state. The strong coupling  regime is needed to exchange this information in a coherent way.

As sketched in Fig. \ref{Fig1}\textbf{d}, a CPW resonator consist of a central planar conductor coupled to external feeding lines through gap capacitors.
The resonant frequency ($f_{\rm CPW}$) depends on the geometrical dimensions of the resonator and can, therefore, be tuned so that $f_{\rm CPW} = f_{\rm G}$.
An electric current flows through the central conductor and the surrounding ground plates with opposite directions.
For a half-wavelength CPW resonator, the produced microwave magnetic field will be maximum at the midpoint of the central conductor, \textit{i.e.}, $z=0$ in Fig. \ref{Fig1}\textbf{d}.
The spatial distribution of this field can be computed numerically (see \textbf{Methods} section for details).
The width and thickness of the central conductor are set to $w = 7$ $\mu$m and $150$ nm, respectively.
The root mean square (rms) of the zero point current fluctuations is given by:
\begin{equation}
\label{irms}
i_{\rm rms} = 2\pi f_{\rm CPW} \sqrt{ \frac{\hbar \pi}{2 Z_0}}.\end{equation}
In this work we consider $Z_0 \sim 50$ $\Omega$ as the inductance of the resonator.\cite{Jenkins2013}
  Eq. \eqref{irms} follows from equating the zero point energy fluctuations to the inductive energy in the resonator, being the only contribution at $z=0$. This yields $\hbar \, 2\pi  f_{\rm CPW}/2 = L I^2/2$. Using the fact that the resonator inductance is $L = Z_0 / \pi^2 f_{\rm CPW}$ we arrive to \eqref{irms}.

Fig. \ref{Fig2}\textbf{a} shows the calculated spatial distribution of the $x$-component of the resulting field $b^x_{\rm rms}(x,y)$ at $z=0$ obtained for ${i_{\rm rms}=11}$ nA. {\color{black} The later corresponds to setting $f_{\rm CPW}=1$ GHz in Eq. \eqref{irms}.}
$\vec b_{\rm rms}$ has only $x$ and $y$ components.
It is maximum at the corners of the central conductor, where the distribution of supercurrent is maximized.
As sketched in Fig. \ref{Fig2}\textbf{a}, we will assume that the magnetic nanodisc is located on top of the central conductor, lying on the $(x,y)$ plane at position $z=x=0$.
Being conservative, we will also assume that the nanodisc bottom edge lies $10$ nm above the upper layer of the superconducting wire.
Therefore, the disc center is positioned at $\textbf{r}_{\rm c} = (0, 10 + r, 0)$ nm where both the $y$ and $z$ components of $b_{\rm rms}$ are negligible, \textit{i.e.}, $ \vec b_{\rm rms} (\textbf{r}_{c}) \approx b^x_{\rm rms} (\textbf{r}_{c}) \vec x$. 
In this way, the zero point rms (in-plane) field can be used to excite the vortex dynamics whereas an external homogeneous (out-of-plane) magnetic field can be used to tune the gyrotropic resonant frequency (see the right inset in Fig. \ref{Fig1}\textbf{c}).

The vortex response $\Delta M_x$ can be increased by increasing the local driving field amplitude $b_{\rm rms}^x(\textbf{r}_{c})$.
This can be easily achieved by decreasing the width of the central conductor (see Fig. \ref{Fig2}\textbf{b}).
For instance, assuming $i_{\rm rms} = 11$ nA and reducing $w$ from $7$ down to $1$ $\mu$m leads to a total increase of $b^x_{\rm rms}$ from $0.6$ up to $4.8$ nT at position $\textbf{r}_{c}$ {\color{black}$=(0,410,0)$ nm}.
Patterning nano-constrictions in localized regions of superconducting CPW resonators has been indeed used as a convenient method to increase the effective coupling between resonant photons and spin qubits.\cite{Niemczyk2010,Jenkins2014}
Reducing $w$ down to, \textit{e.g.}, $\sim 300$ nm leads to a negligible variation in the resulting experimental resonance frequency of the resonator and less than $5 \%$ change in the quality factor.

In Fig. \ref{Fig2}, we see that decreasing $w$ also increases the field inhomogeneity across the disc volume.
This does not affect neither the resulting gyrotropic resonance frequency of the vortex nor the line-width as checked out numerically for different disc sizes. 
As a matter of fact,   we have verified numerically that the relevant parameter, \textit{i.e.}, the disc response $\Delta M_x$ to an oscillating low-amplitude driving field, depends only on the magnetic field at the disc center, \textit{i.e.}, $b_{\rm rms}^x(\textbf{r}_{\rm c})$.


\begin{figure}[t]
\includegraphics[width=0.95\columnwidth]{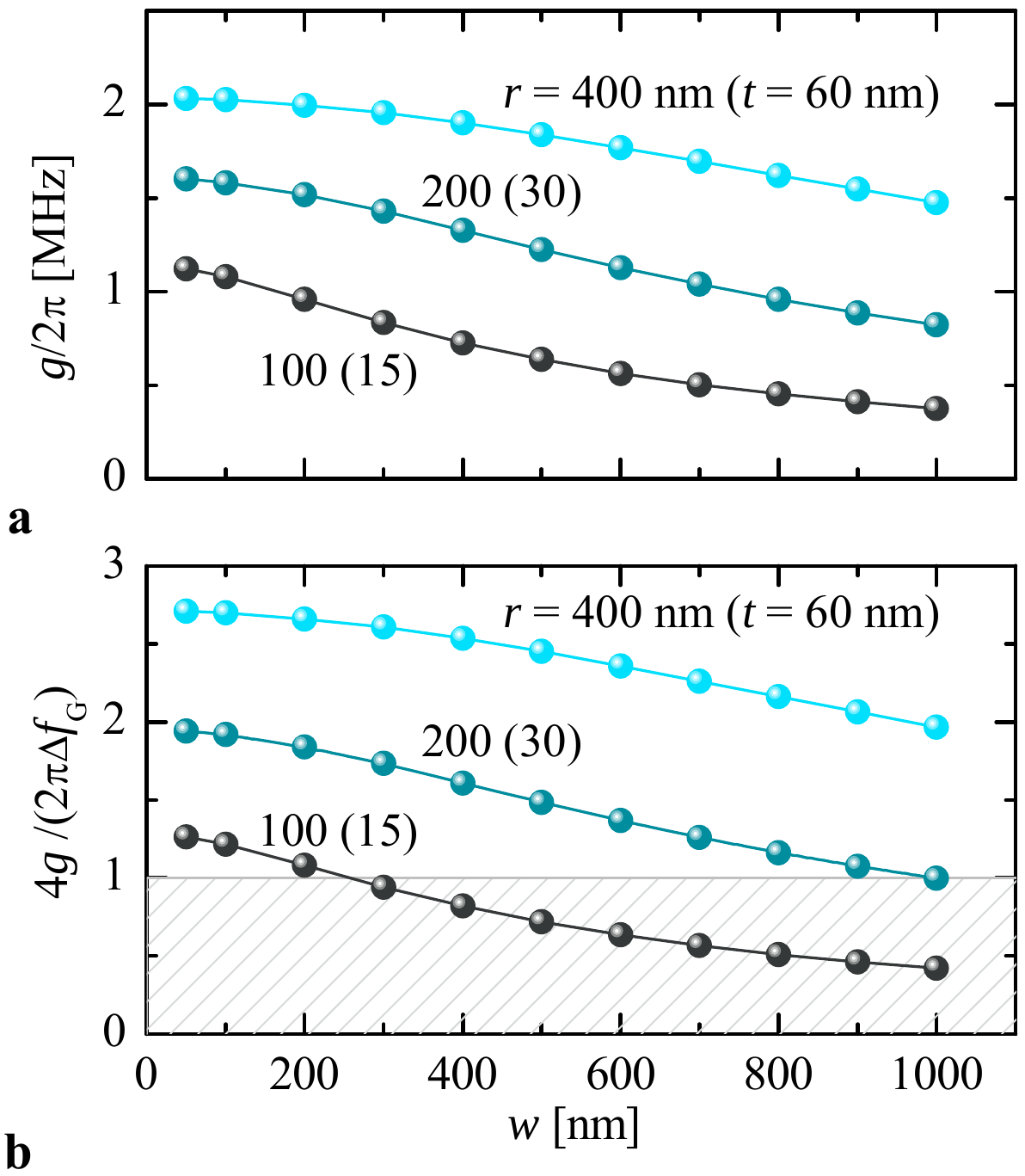}
\caption{\textbf{a}: Numerically calculated coupling $g$ as a function of the central conductor width $w$ for nanodiscs having different sizes. \textbf{b}: Strong coupling condition ($4g/2 \pi \Delta f_{\rm G} > 1$) calculated for the same parameters shown in \textbf{a}.  
}
\label{Fig3}
\end{figure}

\subsection{ Quantization and vortex {\color{black}motion}-photon strong coupling}
Cavity quantization together with the fact that the vortex, around resonance $f_{\rm G}$, can be described using  a harmonic mode, yield  the quantum cavity-vortex model:
{\color{black}
\begin{equation}
\label{H}
H/\hbar = 2 \pi f_{\rm CPW}\;  a^\dagger a +  2 \pi  f_{\rm G} \; a^\dagger_{\rm v} a_{\rm v}
+ 
g  ( a^\dagger a_{\rm v} + {\rm h.c.} )
\end{equation}
}
Here, {\rm h.c.} means hermitean conjugate and $a$ ($a^\dagger$) is a bosonic operator that annihilates (creates) cavity photons with frequency $f_{\rm CPW}$.  The vortex operators  $a_{\rm v}^\dagger$ and $a_{\rm v}$  create and annihilate  single vortex excitations in the gyrotropic mode.  The coupling strength is given by (See {\bf Methods})
\begin{equation}
\label{g}
g = \frac{b_{\rm rms}^x ({\bf r}_{\rm c})}{2} 
\sqrt{\frac{ V \,  \chi_x (f_{\rm G}) \, \Delta f_{\rm G} }{\hbar}} \, ,
\end{equation}
where $V = 2 \pi r^2 t$ is the disc volume. Here, we recall that $b_{\rm rms}^x ({\bf r}_{\rm c})$ is the field generated by the single photon current $i_{\rm rms}$ [Eq. \eqref{irms}] at the center of the vortex, ${\bf r}_{\rm c}$ (Cf. Fig. \ref{Fig2}).
We highlight that Eq. \eqref{g} is based on rather general arguments as the description of the vortex dynamics with a collective variable.\cite{Thiele1973} Therefore, similar arguments  can be extended  to other magnetic topological objects  as skyrmions.

The expression for $g$ can be understood as follows. 
The coupling must be proportional to the magnetic field generated in the resonator. We are interested in the coupling to a single photon, thus, $g \sim b_{\rm rms}$.   The second term in \eqref{g} is analogous to the cavity-magnon coupling case. There, the coupling is proportional to $\sqrt{N}$ with $N$ the number of spins.  This is so since the coupling occurs via the macroscopic spin  operator: $S = S_+ + S_-$ with $S_\pm | l , m \rangle = \sqrt{(l \mp m)(l\pm m +1)} | l , m \pm 1 \rangle $.  Spin-waves are in the maximal spin-state  $l = N/2$ and the spins are in the minimum energy-state $m = \pm N/2$.\cite{Huebl13,SoykalFlatte10} This magnon case, in turn, is understood within the Dicke  model of non-interacting spins which gives the same coupling dependence.  In our discussion, the macroscopic spin  dependence can be understood by noticing that   $\chi \sim M_s / \Delta f_{\rm G}$ and, therefore, $g \sim \sqrt{ V M_s}$.

Numerically, we compute the spatially-dependent $b_{\rm rms}(x,y)$ field and the resulting vortex response around $f_{\rm G}=f_{\rm CPW}$ for different disc sizes with fixed aspect ratio $t/r$. The  gyrotropic frequencies and line-widths obtained from the simulations for each disc are given in Table \ref{tablenew}. {\color{black}In Fig. \ref{Fig3}\textbf{a} we plot the resulting $g$ values obtained when inserting  these parameters into Eq. \eqref{g} vs. the constriction width $w$}. Our numerical simulations show that increasing the volume is beneficial for enhancing the vortex {\color{black} gyration}-cavity coupling. The figure also shows that $g$ saturates when the constriction width is reduced below the disc radius.   Looking at Eq. \eqref{g} we notice that the coupling $g$ is a trade off between the amplitude of the rms field and the disc volume. On the one hand,  $b_{\rm rms}^x $  increases for decreasing radius as  $b_{\rm rms}^x ({\bf r}_{\rm c})$ increases when approaching the superconducting central conductor.  On the other,  the coupling increases with $\sqrt{V}$.

In order to catch these dependences, we approximate \eqref{g} as follows. The numerically calculated field at the disc center is fitted by the simple formula: 
$
b^x_{\rm rms} (\textbf{r}_{\rm c}) = \frac{\mu_0}{2 \pi } i_{\rm rms} \times u_w (r).
$
Here, $u_w(r)$ is a geometrical function of the disc radius depending also on the width of the central conductor. This can be described by $u_w(r) = a_1/r + a_2/r^{\alpha(w)}$ with $0 <\alpha(w) <1$. It satisfies $u_{w} (r) \to 1/r$ for $r/w$ sufficiently large,  \textit{i.e.}, we recover the electrostatic formula of  an infinite line current (see \textbf{Supporting Information}). 
%
Besides, the vortex susceptibility can been approximated   as
$\chi_x = \Delta M_x/ b_{\rm rms}^x ({\bf r}_{\rm c})= (\gamma/2\pi) M_s \xi^2 / \Delta f_{\rm G} $ with $\gamma/2\pi = 28$ GHz/T and $\xi$ a geometrical factor. In the case of discs, $\xi = 2/3$.\cite{Guslienko2006}  
In addition, $f_{\rm G}$ depends on the disc thickness as $t  \cong  9/10 \,  r f_{\rm G} 2 \pi / (\gamma/2\pi) \mu_0 M_{\rm s}$.\cite{Guslienko2006PRL}  
This is only valid for discs having small aspect ratio $t/r \ll 1$ but it will serve to understand the $r$-dependence of the coupling strength. 
Putting altogether we can approximate the coupling in the case of vortices as:
\begin{equation}
\label{gapp}
g \cong \frac{\xi}{4} \sqrt{ \frac{ \pi \mu_0 \, f_{\rm G}^3  }{Z_0}} \; r^{3/2} u_w(r)
\end{equation}
Therefore, fixed the gyrotropic frequency,  the coupling grows with $r$.  The coupling  saturates as $\sim r^{1/2}$ for $w<r$ ($u_w(r) \to 1/r$) in agreement with figure \ref{Fig3}{\textbf a}.

The strong coupling regime occurs when vortex {\color{black} motion} and photons exchange populations coherently in the form of vacuum Rabi oscillations.     The condition to observe such oscillations is  $4 g/ 2 \pi  > \Delta f_{\rm G}$.\cite{Haroche2013} In fig. \ref{Fig3}{\textbf b} we show that the strong coupling regime can be indeed reached using the device sketched in Fig. \ref{Fig1}{\textbf d}.  For radius  $200$ and $400$ nm  strong coupling occurs in the full range of investigated constriction widths. For  smaller discs,  it is reached for constrictions below $300$ nm-width. In the weak coupling regime (shaded region in Fig. \ref{Fig3}\textbf{b}) no coherent exchange excitations occur but an overdamped decay of excitations.  The weak coupling regime is interesting \textit{per s\'e} as it is used to, {\textit e.g.}, control the spontaneous emission of atoms inside cavities. 

\begin{figure}[t]
\includegraphics[width=0.95\columnwidth]{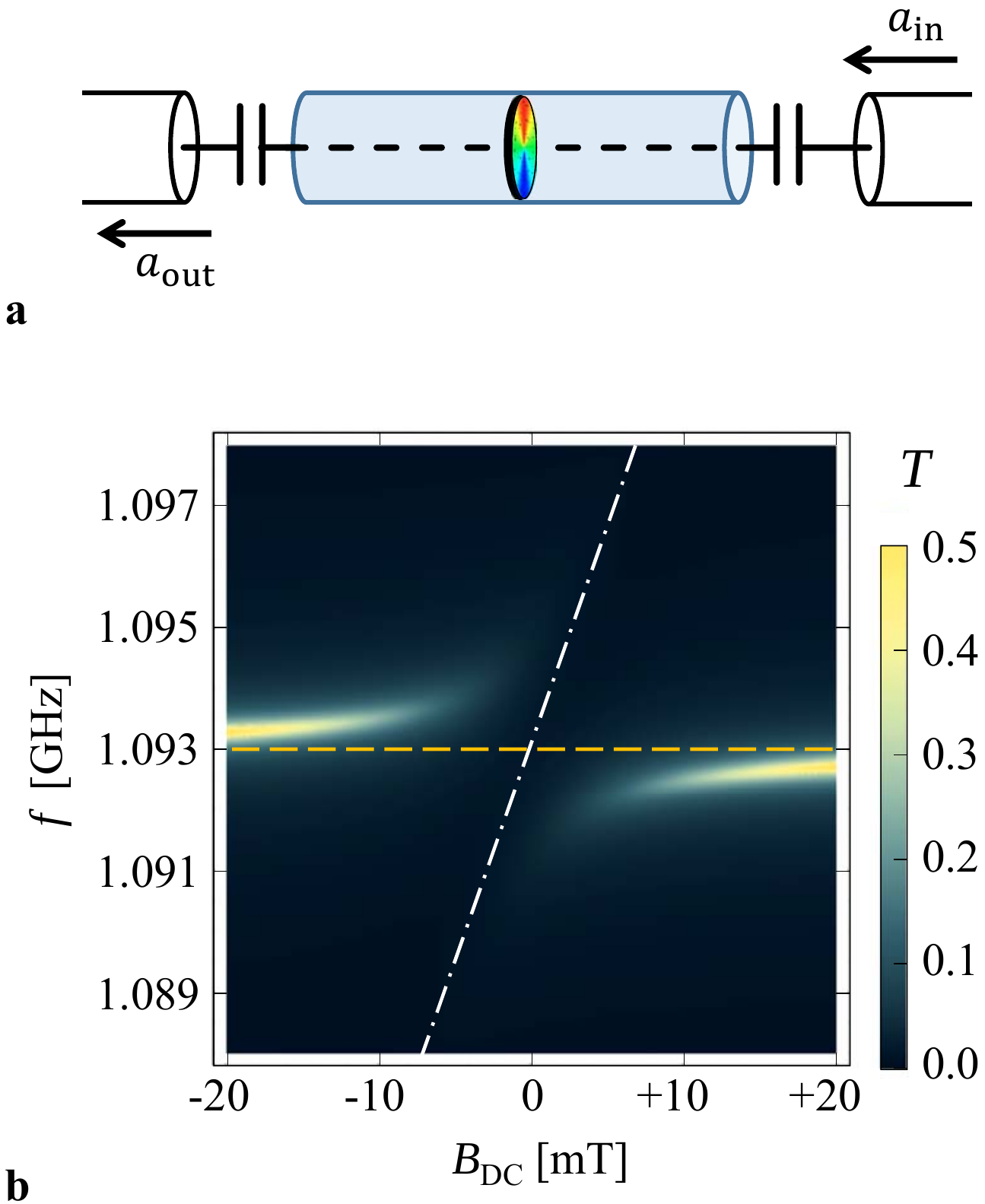}
\caption{\textbf{a}: Schematic representation of a transmission experiment.  The resonator-vortex system is driven through a transmission line using  an input current $a_{\rm in}$.  The output signal $a_{\rm out}$ is then measured.    \textbf{b}:  Transmission $T (f)$ [Eq. \eqref{T}] as a function of the driving frequency $f$ and $B_{\rm DC}$.  The yellow dashed line stands for the resonator frequency $f_{\rm CPW}=1.093$ GHZ while the white dotted-dashed line is the field-dependent vortex frequency $f_{\rm G} (B_{\rm DC})$. Calculations correspond to a disc of  $r=400$ nm and $t= 60$ nm on a $w=500$ nm-wide constriction. 
}
\label{Fig4}
\end{figure}

\subsection{Coupling spectroscopy}
Experimentally, the coupling strength can be measured via a transmission experiment as sketched in Fig. \ref{Fig4}{\textbf a}.
A low-power coherent input signal ($a_{\rm in}$) of frequency $f$ is sent into  the resonator.   The transmitted signal ($a_{\rm out} = T a_{\rm in}$ with $T$ the transmission) is measured with a vector network analyzer.  
The input-output theory\cite{Gardiner1985}  gives the following formula for the transmission:
 \begin{equation}
 \label{T}
 T(f) =
 \left | 
 \frac{\kappa/2}{(f - f_{\rm CPW}) + R + i \Gamma}
 \right |
 \end{equation}
 with $R= \frac{1}{2 \pi}\frac{g^2 (f_{\rm G} - f)}{(f_{\rm G} - f)^2+ \Delta f^2 /4}$ and $\Gamma= \kappa/2 + \frac{1}{2 \pi} \frac{ g^2 \Delta f/2}{(f_{\rm G} - f)^2+ \Delta f^2 /4} $.
$R$ and $\Gamma$ are the real and imaginary part of the self energy of the resonator and $\kappa \sim Q^{-1}$ is the cavity leakage (see \textbf{Supporting Information}). 
Fig. \ref{Fig4}{\textbf{b}} shows the results obtained for a $\textit{P}=+1$ vortex in a $r=400$ nm and $t=60$ nm disc on a $w=500$ nm-wide constriction. The transmission is plotted as a function of the input frequency and the out-of-plane applied magnetic field $B_{\rm DC}$  (see  inset in Fig. \ref{Fig1}{\textbf{c}}). $B_{\rm DC}$ is used to tune the gyrotropic vortex mode on resonance with the resonator frequency.  The latter is chosen to be $f_{\rm CPW} = 1.093$ to coincide with $f_{\rm G}$ at zero $B_{\rm DC}$. At this point, the transmission shows the characteristic double peak.  This is a signature of the strong coupling regime (Cf. Fig. \ref{Fig3}\textbf{b}).   The two peaks are associated to the real part of the zeros of $(f - f_{\rm CPW}) + R + i \Gamma$. At resonance, for $4 g / 2 \pi  > |\Delta f - \kappa |$, the real parts are $f_{\rm CPW} \pm g$, which means that coherent oscillations between a single photon and the vortex {\color{black} motion} can be resolved.  On the other hand, if $ 4 g / 2 \pi < |\Delta f - \kappa |$, the real part of both solutions is  $f_{\rm CPW}$.  In this case, light-matter coupling is not resolved and they are in the weak coupling regime.

\section{Discussion}
We discuss now the feasibility of the proposed experiment.
The fabrication of superconducting CPW resonators with $f_{\rm CPW} \sim 1$ GHz and $Q \sim 10^4 - 10^5$ providing line-widths of few $0.1 - 0.01$ MHz is well within the current state-of-the-art. 
Patterning sub-micron constrictions at the central part of the central conductor poses no technical difficulties neither.
As demonstrated in Ref. \onlinecite{Jenkins2014}, the width of the central conductor can be reduced down to $w \sim 50$ nm by Focused Ion Beam (FIB) milling.
Additionally, applying in-plane external DC magnetic fields of few $\sim$ mT will not affect the properties of resonators based on typical low critical temperature superconductors (\textit{e.g.}, Nb).

Fabrication of an individual sub-micron magnetic disc can be easily achieved by, \textit{e.g.}, conventional electron beam lithography and lift-off or FIB milling of thin magnetic films.
For a given magnetic material, properly setting the geometrical dimensions of the nanodisc will ensure the stabilization of the magnetic vortex.
{\color{black} As discussed previously, ferromagnetic materials with large saturation magnetization are needed. Besides, the aspect ratio shall be made large enough to get $f_{\rm G}\sim$ 1 GHz.  }
%
The vortex line-width, on the other hand, will be given by  
\begin{equation}
\label{linewidth}
\Delta f_{\rm G} = 2 \alpha_{\rm v} f_{\rm G}
\end{equation}
where $\alpha_{\rm v} = \alpha_{\rm LLG} \phi$ is the vortex damping and $\phi = [1+ 1/2\ln(r/r_v)$] is a geometrical factor depending also on the disc thickness through the vortex characteristic radius $r_v$.\cite{Guslienko2006}
%
For the dimensions discussed here, $\phi$ lies within $2 < \phi < 3$.
%
Therefore, we will search for materials exhibiting low Gilbert damping $\alpha_{\rm LLG} \lesssim 10^{-3}$ providing small line-widths of few MHz. 

Finally, a further complication might come from the imposed experimental geometry (see Fig. \ref{Fig1}\textbf{d}). 
Such geometry can be easily achieved in the following way.
The nanodisc can be patterned on a thin (few 100 nm-thick) Si$_3$N$_4$ membrane serving as carrier.
The carrier+nanodisc can be then located perpendicularly to the CPW resonator plane with nanometric resolution using a nanomanipulator in a scanning electron microscope.
Attaching the carrier+nanodisc to the resonator can be easily achieved by FIB induced deposition of Pt.
A similar procedure has been successfully applied to the deposition of micrometric magnetic nanowires over the surface of superconducting sensors achieving $\sim 50$ nm-resolution.\cite{arxiv}

We discuss in the following the convenience of using different ferromagnetic materials (see \textbf{Supporting Information} for further details).
%
For this purpose we compare the strong coupling condition $4g/2\pi \Delta f_{\rm G}$ calculated using Eq. \eqref{gapp} and \eqref{linewidth} for a $r= 400$ nm and $t=60$ nm disc in a $w=500$ nm constriction.
Conventional ferromagnets, \textit{e.g.}, Fe or Ni$_{80}$Fe$_{20}$ (Py), provide large $\mu_0 M_{\rm s} \sim 2.2$ and $\mu_0 M_{\rm s} \sim 1.0$ T, respectively, and intermediate damping characteristics ($\alpha_{\rm LLG} \sim 2 \times 10^{-3}$ and $\alpha_{\rm LLG} \sim 8 \times  10^{-3}$, respectively).\cite{Gladii2017}
These would yield much too low coupling factors with an estimated $4g/2\pi \Delta f_{\rm G} \sim 0.5$ and $0.1$, respectively.
A record low Gilbert damping parameter is exhibited by YIG with $\alpha_{\rm LLG} \sim 5 \times 10^{-5}$ but much too low $\mu_0 M_{\rm s} \sim 0.18$ T leading to $f_{\rm G} \sim 100$ MHz \cite{Yu2014}.
Promising values of  $\alpha_{\rm LLG} \sim 10^{-3}$ have been reported for Heusler alloys like, \textit{e.g.}, NiMnSb ($\mu_0 M_{\rm s} \sim 0.85$ T).\cite{Drrenfeld2015}
However, the latter yields just $4g/2\pi \Delta f_{\rm G} \sim 1$ with the added fabrication complexity of these materials.
More conveniently, CoFe can be easily fabricated by sputtering and offers large saturation magnetization $\mu_0 M_{\rm s} \sim 2.4$ T and low enough Gilbert damping parameters yielding $4g/2\pi \Delta f_{\rm G} \sim 3$.
We highlight that the employed value of $\alpha_{\rm LLG} \sim 5\times 10^{-4}$ was measured experimentally in Ref. \onlinecite{Schoen16} for a 10 nm-thick film at room-temperature. 
%
%
Interestingly, even lower values of $\alpha_{\rm LLG}$ could result from thicker substrates at mK temperatures, yielding larger values of $4g/2\pi \Delta f_{\rm G}$.

\section*{Conclusions}
{\color{black}
Using both numerical and analytical calculations we have shown that strong coupling between the gyrotropic motion of a magnetic vortex in a nanodisc and a single photon in a superconducting resonator is feasible within current technology.
To meet the strong coupling condition we have explored different materials and CPW architectures.  
We have found that CoFe discs with radius $200 < r < 400$ nm and thicknesses $30 < t< 60$ nm are ideal.  Besides, we have argued that patterning sub-micron constrictions  in the middle of the CPW is essential to reach this regime.
Alternatively, the coupling can be further enhanced by using  lower impedance lumped resonators, Cf. Eq. \eqref{gapp}.
%
Finally, our theory can be adapted to other topological excitations as skyrmions. 

Our proposal can be used to transduce photons into quanta of vortex gyration which can be eventually used to emit spin waves.
We notice that the spin waves excited by vortices are sub-micrometer wavelength. This suggest an application  in quantum signal processing.  The reduction of the wavelength traduces in a super slow travel speed  (up to five orders of magnitude slower than light). Therefore, many-wavelength signals could be manipulated on a mm-scale chip.  This is a similar scenario occurring with surface acoustic waves and their coupling to superconducting circuits.\cite{Gustafsson2014,Manenti2017}
In addition, placing more than one disc in the CPW cavity, different gyrating magnetic vortices can be coupled allowing to, {\textit e.g.},  phase-lock distant magnon-emitters or nanoscillators. 

We recall that, both the photons in the resonator and the vortex motion are linear excitations.  
If  we want to exchange single photons or to exploit this device in quantum information protocols nonlinear elements are necessary. 
Circuit QED has already demonstrated strong coupling between a CPW cavity and different types of superconducting circuits.
Thus, we can envision coupling a transmon qubit in one en of the CPW.\cite{Tabuchi15}
This opens the way to further experiments as, {\textit e.g.}, probing vortex-photon entanglement by measuring photon dynamics through qubit dispersive measurements.\cite{Wallraff2005}}
%

{\color{black}
\section{Supporting information
\label{Supporting information}}
Magnetic field generated by the current $i_{\rm rms}$ at the disc center, discussion on the material choice, input-output theory, Hamiltonian diagonalization, normal modes and the rotating wave approximation.
}


\begin{acknowledgements}
We thank Fernando Luis for  inspiring discussions. We acknowledge support by the Spanish Ministerio de Ciencia, Innovaci\'on y Universidades within projects  MAT2015-73914-JIN,   MAT2015-64083-R and MAT2017-88358-C3-1-R, the Arag\'on Government project Q-MAD and EU-QUANTERA project SUMO.
\end{acknowledgements}

\section{Methods}
\label{sec:Methods}

\subsection{Phenomenological quantization}
We discuss here how to obtain the coupling strength $g$ in \eqref{H}.  
Based on our numerical simulations (see main text),  we can approximate  the field fluctuations by their value at the center of the disc: $b_{\rm rms}^x ({\bf r}_{\rm c})$.
The spin-cavity coupling is of the Zeeman type:
\begin{equation}
\label{HZ}
H = \sum_j  \mu_j^x  \,  b_{\rm rms}^x ({\bf r}_{\rm c}) 
\end{equation}
with $\mu_j^x$ the magnetic dipole ($x$-component) of the $j$-spin in the disc. Quantization of the CPW yields:
\begin{equation}
\label{q-B}
\hat b_{\rm rms}  = b_{\rm rms}^x ({\bf r}_{\rm c})  ( a^\dagger + a )
\;.
\end{equation}
The collective variable description for the vortex precession\cite{Thiele1973,Wysin1991,Guslienko2006} allows to write a \textit{quantized} version of the vortex magnetization as:
\begin{equation}
\label{q-M}
\hat \mu_j =  \mu_j (a^\dagger_{\rm v} + a_{\rm v} )
\end{equation}

Using Eqs. \eqref{HZ}, \eqref{q-B} and \eqref{q-M}  we arrive to the coupling constant:
\begin{equation}
\label{hg}
\hbar g =  V \,b_{\rm rms}^x ({\bf r}_{\rm c})   \, \Delta M_x 
\end{equation}
with  $\Delta M_x$ the maximum vortex response in magnetization.  The latter is  obtained by noticing that the vortex itself is driven via the resonator magnetic field with average $\hbar g < a^\dagger + a >$.  In the  single photon limit we can replace $< a^\dagger + a > = 2 \cos (f_{\rm CPW} \tau)$ obtaining:
\begin{equation}
 \Delta M_x  = \frac{4 g}{\Delta f_{\rm G}} b_{\rm rms}^x ({\bf r}_{\rm c})  \;.
\end{equation}
Inserting the above in \eqref{hg} and using $\chi_x (f_{\rm G})= \Delta M_x  / b_{\rm rms}^x ({\bf r}_{\rm c}) $, we end up with  Eq. \eqref{g}.

{\color{black}
Finally, let us comment on the light-vortex motion Hamiltonian, which is a direct consequence of the Zeeman coupling \eqref{HZ}  together with the quantized operators \eqref{q-B} and \eqref{q-M}:
\begin{equation}
\label{Hfull}
H/\hbar = 2 \pi f_{\rm CPW}\;  a^\dagger a +  2 \pi  f_{\rm G} \; a^\dagger_{\rm v} a_{\rm v}
+ g (a + a^\dagger) ( a_{\rm v} + a_{\rm v}^\dagger ) \;.
\end{equation}
For the couplings considered here,  \textit{i.e.}, $g/f_{\rm CPW} \sim 10^{-3}$ (see Fig. \ref{Fig3}), this Hamiltonian can be approximated with the Rotating Wave Approximation (RWA) that dismisses the counter-rotating terms  $a^\dagger a_{\rm v}^\dagger \; + {\rm h.c}$.
In doing so, we arrive to vortex-light Hamiltonian \eqref{H}.
See 
\textbf{Supporting Information} for a discussion on the  validity of the RWA.}

\subsection{Numerical simulations}
Micromagentic simulations are performed using the finite difference micromagnetic simulation package MUMAX3\cite{mumax}.
This software solves the time-dependent Laudau-Lifshitz-Gilbert equation for a given sample geometry and material parameters assuming zero temperature.
The saturation magnetization, exchange stiffness constant and Gilbert damping are set to $M_{\rm s} = 1.9 \times 10^6$ A/m, $A = 2.6 \times 10^{-11}$ J/m and $\alpha_{\rm LLG} \sim 5 \times 10^{-4}$ for CoFe, respectively.
We use a box with dimensions $2r \times 2r \times t$ containing a disc of radius $r$ and thickness $t$. 
For the $r=100$ nm ($t = 15$ nm), $200$ (30) and $400$ (60) discs, the boxes are discretized into $64 \times 64 \times 8$, $128 \times 128 \times 16$ and $256  \times 256  \times 32$ identical cells, respectively, each being $3.12 \times 3.12 \times 1.87$ nm$^3$. 
The vortex ground-energy configuration of the disc is found by relaxing the system in the presence of an homogeneous external out-of-plane magnetic field ($B_{\rm DC}$).

Vortex dynamics are characterized using an in-plane perturbation magnetic field $b(x,y,t) = b_{x,y} b_t$ where we have splitted the   space-  and time-varying parts.
$b_{x,y}$ is created by the current $i$ flowing through the central conductor of the CPW resonator being, therefore, non-homogeneous along the $(x,y)$ plane and having a negligible $z$ component.
$b_{x,y}$ is calculated numerically using 3D-MLSI\cite{Khapaev2002} which allows to obtain the spatial distribution of supercurrents circulating through 2-dimensional sheets in a superconducting wire.
The input parameters are the current $i$ flowing through the central conductor having width $w$ , thickness 150 nm and London penetration depth  $\lambda_{\rm L} = 90$ nm (typical for Nb).

The gyrotropic mode is characterized using $i=10$ mA and $b_t = {\rm sinc}  (2\pi f_{\rm cutoff} t)$ and calculating numerically the FFT of the resulting time-dependent spatially-averaged magnetization along the $x$ direction $M_x(t)$.
This allows us to obtain the gyrotropic characteristic frequency ($f_{\rm G}$) and line-width ($\Delta f_{\rm G}$) by fitting to a Lorentzian curve $$ L(f) = \frac{A}{\pi}\frac{(\Delta f_{\rm G})/2}{(f- f_{\rm G})^2+(\Delta f_{\rm G}/2)^2}. $$

The vortex  response  to quantum vacuum fluctuations is found by using $b_{x,y}=b_{\rm rms}(x,y)$ and $b_t = \cos (2\pi f_{\rm G} t)$ where $b_{\rm rms}(x,y)$ is the field resulting when $i=i_{\rm rms}$. 
The latter is obtained from Eq. \eqref{irms} for $f_{\rm CPW}=f_{\rm G}$ giving  $i_{\rm rms} \sim 10 - 20$ nA for the resonator parameters used here.
In this way, $\chi_x(f_{\rm G}) =\Delta M_x / b^x_{\rm rms}$ where $\Delta M_x$ corresponds to $M_x(t) = \Delta M_x \sin(2\pi f_{\rm G} t) $ at resonance.


\bibliography{vortex}


\end{document}